# Sufficient Conditions for Observability in Electrocardiography Problem

**Raheam. A. Mansor Al-Saphory**
*Department of Mathematics, College of Education for Pure Science, Tikrit University, Tikrit , Iraq*
**E-mail:** saphory@hotmail.com

## Abstract:
The aim of this paper, is to study electrical cardiography problem. Thus, we construct the state space system of this model as mathematical model. Moreover, we present some definitions and results which is described some concepts of linear control system analysis related to this problem. More precisely, the sufficient conditions which characterize the observability notion of linear dynamical controlled system are presented and discussed. Finally, we prove that, the electrical cardiography model is completely observable system over finite time $t\epsilon\,[0,T]$.

**Key words:** Cardiography model, Observable systems, Linear system, Dynamical system.

## 1. Introduction
The great progress in control science since 1955 has change basic concept of analysis and synthesis of control system. This progress has depended largely on mathematical study of optimal control systems[1]. Modern control theory which is based on state space concepts is extremely useful not only for designing a specific optimal control system but also for improving the principle on which the system well operate [2]. In recent years, control system have assumed an increasing important role in the development and advancement of modern civilization and technology. Practically every aspect of our day-to-day activities is affected by some type of control system. Control system are found in abundance in all sectors of industry, such machine-tool control, quality of control manufactured products, automatic assembly line, ... [3]. Considered system may be described by the following linear dynamical form

$$\begin{cases} \dot{x}(t) = Ax(t) + Bu(t) \\ x(0) = x_0 \\ x(T) = 0 \end{cases} \qquad (S_1)$$

where $A$, $B$ are $n\times n$ and $n\times p$ matrices (respectively), $x(t)\epsilon\,L^2(0,T;R^n)$ is the Hilbert state space with $x\epsilon\,R^n$, $u(t)\epsilon\,L^2(0,T;R^p)$ is the Hilbert control space with $u\epsilon\,R^p$, and $\dot{x}(t) = Ax(t) + Bu(t)$ is the state space equation with initial state $x_0\,\epsilon\,L^2(0,T;R^n)$ and final state $x(T)\epsilon\,L^2(0,T;R^n)$. The system is augmented by the following output function

$$y(t) = Cx(t)$$

$(S_2)$ Where $C$ is $n\times$ and $y(t)\epsilon\,L^2(0,T;R^q)$ is the Hilbert observation space with $y\epsilon\,R^q$. The systems $(S_1)$-$(S_2)$ are more general mathematical model represent various cases [1-3].

The problem of feedback control, it is common to think of biological systems as fragile. However, most are very stable, and it is almost a tautology to say so, because they must all operate in the fact of changing and fluctuating environmental parameters; so if they weren't stable, they wouldn't be here. We are familiar from engineering with the concept of feedback control whereby variables sensed and parameters are then rest to change the behavior of the system. The nephrons in the kidney sense ***Nacl*** concentration in the blood and adjust filtration rate to regulate salt and water in the body. The baroreceptor loop regulate blood pressure, heart rate, and peripheral resistance to adjust the circulation to different challenge. Numerous such control systems are known and studied in animal and plant physiology [4].

Mathematical modeling of blood flow and electric heart activity have been researched extensively throughout the previous decades.

There are multiple reasons for this focus. Firstly, cardiovascular diseases are leading cause of death in the developed part of world. Secondly, heart activity and arterial blood flow can appropriately described by equations already known from physics. Consequently, the major research effort is now on the design of efficient computer programs for obtaining accurate approximations [5, 6].

For the untrained in mathematics, it can be unclear exactly what a mathematics model is. In figure 1, an approximate solution to a mathematical model of blood flow is shown.

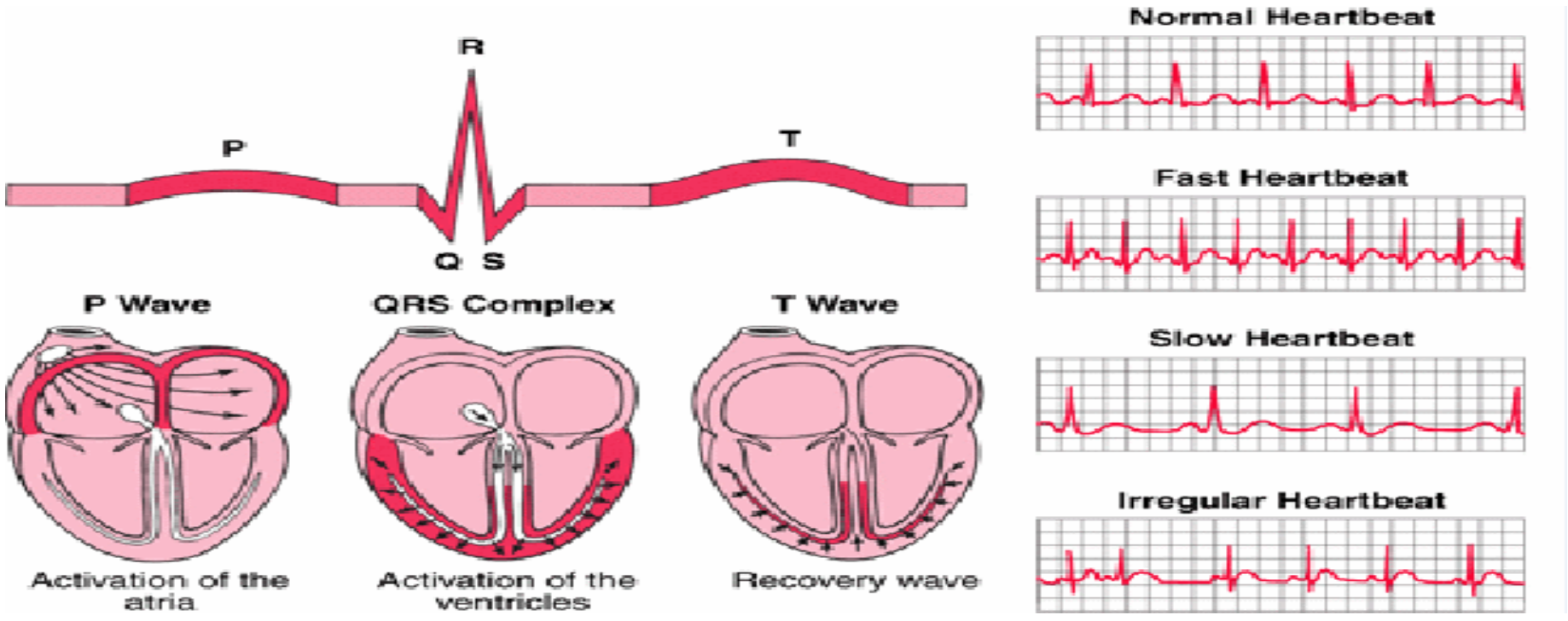

**Fig. 1: ECG of various electrocardiography with deferent heartbeat cases.**

It is a snapshot of the flow at a given time angle. Observe that only the flow in a slice of artery is shown. The mathematical model, and the algorithm used to approximate a solution, independent of such choices (time, angle, slice, etc,…) [7]. Thus the model in figure 1 is a spatial case of general mathematical model systems $(S_1)$- $(S_2)$. Thus, the same computer program can be used for any set of choice. Consequently, very detailed studies of the flow can be performed. Moreover, the model does not depend on the actual geometry of the artery. Thus, the same computer program can be used on any artery and any patient. One can even apply modifications to an artery and see how this change the flow. In this way, it might be possible to predict the outcome of a surgery without actually having to perform it. For more dissection, see [4]. For more detailed exposition of the mathematics and numerical solution approaches, see [8].

The purpose of this work is to study the electrical cardiography model and to prove that, this model is completely observable system through state space analysis. The outline of this, is organized as the following:

Section 1, concerns some definitions and characterizations in control systems. Section 2, related to study the solution method of linear control system and some mathematical approaches. Later section 3, devotes the observability notion of electrical cardiography model as control system.

## 2. Some definitions and characterizations in control systems

In this section, we present some preliminaries related to the state space analysis as in ref.s $[9-10]$ and we give some definitions and characterizations concern linear dynamical control systems.

**Definition 2.1:**

State space analysis is very useful technique of analyzing control system. It is based on the concept of state and is applicable to linear time varying, non-linear and multi-input multi-output systems. Thus, representation of higher order system become simple.

**Definition 2.2:**

In general, differential equation of an $n$th-order system is written by

$$y^n(t) + a_n y^{n-1}(t) + \ldots + a_2\dot{y}(t) + a_1 y(t) = F(t) \quad (1)$$

Which also known as a linear ordinary differential equation if the coefficients $a_n, a_{n-1}, \ldots, a_1$ are not functions of $y(t)$. In this paper, because we treat only systems that contain lumped parameters, the differential equations encountered are all of the ordinary type [3]. For the systems with distributed parameters, such as in heat-transfer systems, partial differential equations are used $[3, 11-12]$.

**Remark 2.3:**

Let us define

$$\begin{cases} x_1(t) = y(t) \\ x_2(t) = \dot{y}(t) \\ \quad \vdots \\ x_n(t) = y^{(n-1)}(t) \end{cases}$$

An $n$th-order differential equation can be decomposed into $n$ first-order differential equations as following

$$\begin{cases} \dot{x}_1(t) = x_2(t) \\ \dot{x}_2(t) = x_3(t) \\ \quad \vdots \\ \dot{x}_{n-1}(t) = x_n(t) \end{cases} \quad (2)$$

From equations (1) and (2), we have

$$\begin{cases} \dot{x}_n(t) = y^n(t) = \\ F(t) - a_n x_n(t) \ldots - a_1 x_1(t) = \\ -a_1 x_1(t), - \cdots -, a_n x_n(t) + F(t) \end{cases} \quad (3)$$

Because, in principle first-order differential equations are used in the analytical studies of control systems. Notice that, the last equation (3) is obtained by the highest-order derivative term in equation (1) to the rest of the terms. In control systems theory, the set of first order differential equations in (3) is called the state equations, and $x_1, x_2, \ldots, x_n$ are called the state variables.

**Remark 2.4:**

The state of a system refers to the past, present and future conditions of the system from mathematical perspective, it is convenient to define a set of a state variables and state equations to model dynamic systems. As it turns out, the variables

$$x_1(t), x_2(t), \ldots, x_n(t) \quad (4)$$

defined in equation (2.2) are the state variable of $n$th-order system described by (1), and the $n$th-order differential equations are the state equations.

In general, there are some basic rules regarding the definition of a state and what constitutes a state equation. The state variables must satisfy the following conditions:

$*$ At any time initial $t = t_0$, the state variables

$$x_1(t_0), x_2(t_0), \ldots, x_n(t_0) \quad (5)$$

define the initial states of the system.

$*$ Once the inputs of the system for $t \geq t_0$ and initial states just defined are specified, the state variables should completely define the future behavior of the system.

The state variables of a system are defined as a minimal set of variables,

$$x_1(t), x_2(t), \ldots, x_n(t)$$

Such that the knowledge of these variables at any time $t_0$ and information on the applied input at time $t_0$ are sufficient to determine the state of the system at any time $t > t_0$. Hence, the space state form for $n$ variables is given by

$$\dot{x}(t) = Ax(t) + Bu(t) \quad (6)$$

Where $x(t)$ is the state vector having $n$ rows,

$$x(t) = \begin{bmatrix} x_1(t) \\ x_2(t) \\ \vdots \\ x_n(t) \end{bmatrix}$$

And $u(t)$ is the input vector with $p$ rows,

$$u(t) = \begin{bmatrix} u_1(t) \\ u_2(t) \\ \vdots \\ u_p(t) \end{bmatrix}$$

The coefficient matrices $A$ and $B$ are defined by

$$A = \begin{bmatrix} a_{11} & \cdots & a_{1n} \\ \vdots & \ddots & \vdots \\ a_{n1} & \cdots & a_{nn} \end{bmatrix} \quad (7)$$

and

$$B = \begin{bmatrix} u_{11} & \cdots & u_{1n} \\ \vdots & \ddots & \vdots \\ u_{n1} & \cdots & u_{np} \end{bmatrix} \quad (8)$$

**Definition 2.5:**

An output of a system is a variable that can be measured, but state variable does not always satisfy this requirement. For instance, in an electric motor, such variables as winding current, rotor velocity, and displacement can be measured physically, and these variables all qualify as output variables. In general, output can expressed as an algebraic combination of the state variables. For the system described by equation (1), if $y(t)$ is designed as the output equation (function) is simply given by

$$y(t) = x_1(t)$$

then, in general, we have

$$y(t) = \begin{bmatrix} y_1(t) \\ y_2(t) \\ \vdots \\ y_n(t) \end{bmatrix} = Cx(t) \quad (9)$$

where

$$C = \begin{bmatrix} C_{11} & \cdots & C_{1n} \\ \vdots & \ddots & \vdots \\ C_{q1} & \cdots & C_{qn} \end{bmatrix} \quad (10)$$

**Definition 2.6:**

State space is the $n$-dimensional space coordinates axis consists

$$x_1 - axis, x_2 - axis, \dots, x_n - axis$$

Any state can uniquely represented by a point in the state spaces.

**Definition 2.7:**

Consider the following differentiable equation

$$\dot{x}(t) = f\big(t, x(t), u(t)\big), -\infty < t < \infty \quad (11)$$

This equation equivalent the set of $n$ scalar differentiable- equation

$$\dot{x}(t) = \frac{dx(t)}{dt} = f_i(t, x_1(t), \dots, x_n(t), u_1(t), \dots, u_p(t)) \quad (12)$$

where $i = 1, 2, \dots, n$. The $i$th state variable is represented by $x_i(t)$ and $u_j(t)$ denotes the $j$th input for $j = 1, 2, \dots, p$, is called dynamical system, where $x(t) \in R^n$ is a state vector and $u(t) \in R^p$ is control vector and $t \in [0, T] \subseteq R$ is the time and then

$$f : R \times R^n \times R^p \to R^n$$

and

$$f \in C^1(D),$$

Where

$$D \subseteq R \times R^n \times R^p$$

where $D$ is the domain of the function $f$. For ease of expression and manipulation it is convenient to represent the dynamical system equations in vector-matrix from. Let us define the following vectors:

*State vector:*

$$x(t) = \begin{bmatrix} x_1(t) \\ x_2(t) \\ \vdots \\ x_n(t) \end{bmatrix}$$

*Input vector:*

$$u(t) = \begin{bmatrix} u_1(t) \\ u_2(t) \\ \vdots \\ u_p(t) \end{bmatrix}$$

*Output vector:*

$$y(t) = \begin{bmatrix} y_1(t) \\ y_2(t) \\ \vdots \\ y_{q(t)} \end{bmatrix}$$

By using these vector, the $n$ state equations of equation (12) can be written as

$$\dot{x}(t) = f\big(t, x(t), u(t)\big) \quad (13)$$

where $f$ denotes an $n \times 1$ column matrix that contains the functions $f_1, f_2, \dots, f_n$ as elements. Similarly, the $q$ output functions in (9) is given by

$$y(t) = Cx(t) \quad (14)$$

where $C$ denotes a $q \times 1$ column matrix that contains functions $C_1, C_2, \dots, C_q$ as elements.

**Definition 2.8:**

The dynamical controlled system

$$\dot{x}(t) = f(t, x(t), u(t))$$

Is called free (unforced) dynamical system if

$$u(t) \cong 0, \forall t \in [t_0, t_1] \subseteq R$$

This system can be written as follows:-

$$\dot{x}(t) = f(t, x(t))$$

(15) **Definition 2.9:**

The dynamical forced system (2.13)

$$\dot{x}(t) = f(t, x(t), u(t))$$

is stationary if, we have

$$f\big(t, x(t), u(t)\big) = f\big(x, u(t)\big) \quad (16)$$

then, for all $t \geq 0$,

$$\dot{x}(t) = f(x, u(t)) \quad (17)$$

i.e., the function $f$ depend implicitly on time $t$ through $u(t)$.

**Definition 2.10:**

The dynamical controlled system

$$\dot{x}(t) = f(t, x(t), u(t))$$

is called autonomous dynamical system, if satisfies the following conditions:

1- free
2- stationary

thus, the system (13) is given by the form

$$\dot{x}(t) = f(x)$$

**Definition 2.11:**

The dynamical controlled system

$$\dot{x}(t) = f\big(t, x(t), u(t)\big)$$

is linear if $f$ is linear function of $x$ and $u$. This system is given by the following equation

$$\dot{x}(t) = A(t)x(t) + B(t)u(t) \quad (18)$$

Where

$$A(t) = \left(\frac{\partial f_i}{\partial x_i}\right)_{n\times n} = \begin{bmatrix} \frac{\partial f_1}{\partial x_1} & \cdots & \frac{\partial f_1}{\partial x_n} \\ \vdots & \ddots & \vdots \\ \frac{\partial f_n}{\partial x_1} & \cdots & \frac{\partial f_n}{\partial x_n} \end{bmatrix} \quad (19)$$

And

$$B(t) = \left(\frac{\partial f_i}{\partial u_j}\right)_{n\times p} = \begin{bmatrix} \frac{\partial f_1}{\partial u_1} & \cdots & \frac{\partial f_1}{\partial u_p} \\ \vdots & \ddots & \vdots \\ \frac{\partial f_n}{\partial u_1} & \cdots & \frac{\partial f_n}{\partial u_p} \end{bmatrix} \quad (20)$$

Where $i = 1,\dots,n, j = 1,\dots,p$ and $A, B$ respectively are $n \times n$, $n \times m$ matrices depending on time $t$. Then, the system (18) is called linear time varying (continuous) controlled system.

**Definition 2.12:**

The linear time varying dynamical controlled system

$$\dot{x}(t) = A(t)x(t) + B(t)u(t)$$

Is called free (unforced) linear dynamical system if

$$u(t) \cong 0, \forall\, t \in [t_0, t_1] \subseteq R$$

This system can be written as follows:-

$$\dot{x}(t) = A(t)x(t) \quad (21)$$

**Definition 2.13:**

The linear time varying dynamical controlled system

$$\dot{x}(t) = A(t)x(t) + B(t)u(t)$$

is stationary if, we have

$$A(t)x(t) + B(t)u(t) = Ax(t) + Bu(t) \quad (22)$$

then for all $t \geq 0$, we obtain

$$\dot{x}(t) = Ax(t) + Bu(t) \quad (23)$$

i.e., the function $f$ depend implicitly on time $t$ through $u(t)$.

**Definition 2.14:**

The linear time varying dynamical controlled system (18) is called autonomous linear dynamical system, if it is:

1- free
2- stationary

Thus, the system (18) is given by

$$\dot{x}(t) = Ax(t) \quad (24)$$

**Definition 2.15:**

Consider linear dynamical controlled system with initial state and final state described by the following state space equations

$$\begin{cases} \dot{x}(t) = Ax(t) + Bu(t) \\ x(0) = x_0 \\ x(T) = 0 \end{cases} \quad (25)$$

augmented with the output function

$$y(t) = Cx(t) \quad (26)$$

Where $C$ is $q \times n$ matrix. The systems $(25) - (26)$ are said to be observable, if for unknown initial state, there exists a finite $t \geq 0$ such that the knowledge the input $u(t)$ and the output $y(t)$ over $[0, T]$ suffices to determine uniquely, the initial state $x(0)$.

Otherwise the systems $(25) - (26)$ are called un observable system.

**Definition 2.16:**

The systems $(25) - (26)$ are completely observable system, if for every initial state $x_0$, there exists time $t \geq 0$ such that, the knowledge of the input $u(t)$ and the output $y(t)$ suffices to determine uniquely, the initial state $x_0$.

**Definition 2.17:**

For linear dynamical systems $(25) - (26)$, the observability matrix $M(0,T)$ is defined by the following formula

$$M(0,T) = \int_0^T e^{A^*(T-\tau)} C^* C e^{A(T-\tau)} \quad (27)$$

Where $e^{A(T-\tau)}$ is $n \times n$ matrix and, $C$ is $n \times m, C^*$ is the conjugate transpose of $B$ and $e^{A^*(T-\tau)}$ is the conjugate transpose of $e^{A(T-\tau)}$.

**Definition 2.18:**

The matrix $M(0,T)$ is called positive definite if

$$< M(0,T)x, x >> 0, \forall\, x \in R^n\ x \neq 0 \quad (28)$$

and is called positive semi-definite if

$$< M(0,T)x, x > \geq 0, \forall\, x \in R^n \quad (29)$$

i.e., $\exists\, x \neq 0$, such that

$$< M(0,T)x, x > = 0$$

## 2.2. State transition matrix

This sub-section related to recall some definition and characterization as in [5].

**Definition 2.19:**

The state transition matrix is defined as matrix that satisfied the linear free dynamical system.

$$\dot{x}(t) = Ax(t) \quad (30)$$

If $\Phi(t)$ be $n \times n$ matrix that represents the state transition matrix, then

$$\dot{\Phi}(t) = A\Phi(t) \quad (31)$$

Let $x(0)$ is the initial state at $t = 0$. Then $\Phi(t)$ also defined by the matrix equation

$$x(t) = \Phi(t)x(0) \quad (32)$$

Which is the solution of the free linear dynamical system (31) for $t \geq 0$.

**Remark 2.20:**

To find state transition matrix, we use (Laplace transform approach) to find $\Phi(t)$, we take Laplace transform on both sides of system (30), we have

$$SX(s) - x_0 = AX(s)$$

thus, we obtain

$$SX(s) - AX(s) = x_0$$

and then

$$X(s)[SI - A] = x_0$$

therefore, we can get

$$L(x(s)) = ((SI - A)^{-1})\, x(0)$$

Where assumed that matrix $(SI - A)$ is non-singular (That means $det.(SI - A) \neq 0$). By taking the inverse of Laplace transform on both sides of the equation

$$L^{-1}L(x(s)) = L^{-1}((SI - A)^{-1})\, x(0)$$

we have,

$$x(t) = L^{-1}((SI - A)^{-1})\, x(0),\ t \geq 0 \quad (33)$$

Comparing equation (32) with equation (33), the state transition matrix is defined by

$$\Phi(t) = L^{-1}((SI - A)^{-1} = e^{At} \quad (34)$$

then, we have from eq. (33) and eq. (34), the solution of linear free dynamical system (30) given by the following formula [5]:

$$x(t) = \Phi\,(t)x(0) = e^{At}\,x(0) \qquad (35)$$

## 3. Mathematical Method

In this section, we discuss the solution method of linear control system and some mathematical approaches as in ref.s [9 − 10].

**3.1. The method of solution**

Consider the system described by the state space equation

$$\dot{x}(t) = Ax(t) + Bu(t) \qquad (36)$$

Augmented with output function

$$y(t) = Cx(t) \qquad (37)$$

Where $A, B$ and $C$ are respectively $n \times n$ , $n \times m$ and $q \times n$ constant matrix. The problem is to find the solution excited by initial state $x(0)$ and the input $u(t)$. Thus, the solution hinges on the exponential function of $A$, we need

$$\frac{d}{dt}\,e^{At} = A\,e^{At} = e^{At}A$$

To develop the solution of system (36), then, we multiply (36) by $e^{-At}$, we have

$$e^{-At}\,\dot{x}\,(t) - e^{-At}Ax\,(t) = e^{-At}\,Bu\,(t)$$

this implies

$$\frac{d}{dt}\left(e^{-At}\;x\,(t)\right) = e^{-At}\;Bu\,(t)$$

By integration the above equation from 0 to $t$ yields

$$e^{-At}\;x\,(t) - x\,(0) = \int_0^t e^{A\tau}\;Bu\,(\tau)\,d\,\tau$$

Because the inverse of $e^{-At}$ is $e^{At}$ and $e^{0} = I$, then, we have

$$x\,(t) = e^{At}x(0) + \int_0^t e^{A(T-\tau)}\;Bu\,(\tau)\,d\,\tau$$

(38)

Therefore, $x(t)$ in equation (38) is the solution of the system (36) (see ref. [10]).

**Remark 3.1:**

The systems (36) − (37) is completely observable if

$$\forall\, x_0 \neq x_1 \in R^n$$

Initial states imply that, the output functions

$$y_0(t) \neq y_1(t)$$

**3.2. Characterization of observable system**

The observability notion of the linear dynamical controlled system in ref.s [9, 10] can be developed in a new way by the following result:

**Theorem 3.2:**

The linear dynamical system

$$\begin{cases} \dot{x}(t) = Ax(t) + Bu(t) \\ x(0) = x_0 \\ x(T) = 0 \end{cases} \qquad (39)$$

with output function

$$y(t) = Cx(t) \qquad (40)$$

is completely observable to zero over $[0, T]$ … (1)

⇔ The observability matrix $M(0, T)$ is invertible ... (2)

⇔ The observability matrix $M(0, T)$ is positive definite … (3)

Now, we prove this theorem by the following way

(3) ⇒ (2) ⇒ (1) ⇒ (3)

for achieve the observability of cardiography model.

**Proof:**

(3) ⇒ (2)

If the observability matrix $M(0, T)$ is positive definite, to prove that $M(0, T)$, is invertible. Now, if $M(0, T)$ is positive definite, that means

$$< M(0, T)x_0, x_0 >> 0, \forall\, x_0 \neq 0$$

and

$$< M(0, T)x_0, x_0 >> 0, \qquad \text{if } x_0 = 0$$

Since the matrix $M(0, T)$ is positive definite, then, the matrix $M(0, T)$ has no zeros eigenvalues, and if $M(0, T)$ has no zeros eigenvalues, that means, the determinant of $M(0, T) \neq 0$, Therefore, $M(0, T)$ is invertible [9].

**Proof :**

(2) ⇒ (1)

If $M(0, T)$ is invertible, to prove that, the system (39) together with output function (40)

$$\begin{cases} \dot{x}(t) = Ax(t) + Bu(t) \\ x(0) = x(0) \\ x(T) = 0 \\ y(t) = Cx(t) \end{cases} \qquad (41)$$

is completely observable over $[0, T]$. We know that, the solution of the linear free dynamical system (39)

$$\dot{x}(t) = Ax(t)$$

is given by

$$x(t) = \Phi\,(t)x_0 = e^{At}\,x_0$$

and

$$y(t) = Cx(t) = Ce^{At}\,x_0$$

Now, we can calculate

$$M(0, T)x_0 = \int_0^T e^{A^*(T-\tau)}C^*Ce^{A(T-\tau)}\,x_0 d\tau$$

and since, the observability matrix $M(0, T)$ is invertible, then, we can evaluate the initial state by the following form

$$x_0 = M(0, T)^{-1}\int_0^T e^{A^*(T-\tau)}C^*Cy(\tau)\,d\tau$$

and if, we choose

$$x_0 \neq x_1$$

then, we have

$$y_0(t) \neq y_1(t)$$

where the output functions $y_0(t)$, $y_1(t)$ are given by:

$$y_0(t) = Ce^{At}\,x_0$$

and

$$y_1(t) = Ce^{At}\,x_1$$

Then by remark (3.1), the linear dynamical system (41) is completely observable over $[0, T]$.

**Proof:**

(3)⇒(1)

If the system (41) is completely observable over $[0, T]$, to prove that $M(0, T)$ is positive definite. We can calculate

$$< M(0, T)x_0, x_0 > =$$

$$< \int_0^T e^{A^*(T-\tau)}\,C^*Ce^{A(T-\tau)}d\tau\; x_0, x_0 >$$

$$= \int_0^T < e^{A^*(T-\tau)} C^* C e^{A(T-\tau)} x_0, x_0 > d\tau$$

$$= \int_0^T < C e^{A(T-\tau)} x_0, C e^{A(T-\tau)} x_0 > d\tau$$

$$= \int_0^T < y_0(\tau), y_0(\tau) > d\tau$$

$$= \int_0^T \|y_0(\tau)\|^2 d\tau \geq 0$$

$\Rightarrow M(0,T)$ is positive semi definite.

Since the system (41) is completely observable over $[0,T]$, then

$$\forall x_0 \neq x_1 \in R^n, \exists u : [0,T] \to R^n$$

such that

$$y_0(t) \neq y_1(t)$$

That means

$$\exists x_0 \neq 0 \text{ implies } y_0(t) \neq 0,$$

and hence

$$< M(0,T)x_0, x_0 >= \int_0^T \|y_0(\tau)\|^2 d\tau > 0$$

Finally, $M(0,T)$ is positive definite ■.

The sufficient condition to characterize observable system is given by the following result:

**Theorem 3.3:**

The linear controlled system (41)

$$\begin{cases} \dot{x}(t) = Ax(t) + Bu(t) \\ x(0) = x(0) \\ x(T) = 0 \\ y(t) = Cx(t) \end{cases}$$

is completely observable over $[0,T]$, if the

$$rank\left(C^*, A^*C^*, \quad A^{*^2}C^*, \dots, A^{(n-1)}C^*\right) = n$$

**Proof:**

If the rank of the following matrix

$$\left(C^*, A^*C^*, A^{*^2}C^*, \dots, A^{(n-1)}C^*\right) = n,$$

to prove that, the system (41) is completely observable over $[0,T]$. Now , if the system (41) is not observable over $[0,T]$. That is means, the observability matrix $M(0,T)$ is not positive definite. By using the previous theorem 3.2, and if we choose $x_0 \neq 0$, then, we have

$$< M(0,T)x_0, x_0 > = \int_0^T \left\|Ce^{A(T-\tau)}x_0\right\|^2 d\tau = 0$$

thus, put $s = T - \tau$, implies that

$$Ce^{A(s)}x_0 = 0, \quad \forall s \in [0,T]$$

By deriving the above equation multiple once, we have

$$CAe^{As}x_0 = 0, \quad \forall s \in [0,T]$$
$$CA^2 e^{As}x_0 = 0, \quad \forall s \in [0,T]$$
$$\vdots$$
$$CA^{(n-1)} e^{As}x_0 = 0, \quad \forall s \in [0,T]$$

Put $s = 0$ , we have

$$CI_n x_0 = 0 \text{ implies } \quad x_0{}^* C^* = 0$$

by multiply both sides of above equation $A$, we obtain

$$CAx_0 = 0 \text{ implies } \quad x_0{}^* A^* C^* = 0$$
$$\vdots$$
$$CA^{(n-1)}x_0 = 0 \text{ implies } \quad x_0{}^* A^{*(n-1)} C^* = 0$$
$$\Rightarrow rank\left(C^*, A^*C^*, A^{*^2}C^*, \dots, A^{(n-1)}C^*\right) = 0$$

This is (Contradiction), Because,

$$rank\left(C^*, A^*C^*, A^{*^2}C^*, \dots, A^{(n-1)}C^*\right) = n$$

Therefore, the system (41) is completely observable over $[0,T]$ ■.

## 4. Application to Cardiography Model

In this section, we give a physical model as dynamical system and by using state space analysis transform this model to linear control observable system.

### 4.1. The physical model

The field of medicine that deals with study of heart is called cardiology as in ref. [13], the nature and effects of vibrations of the heart as pumps blood through the circulatory system of body are a great source of mathematical application as. An important aspect involves the recording of such vibrations known as cardiography model as in figure 2.

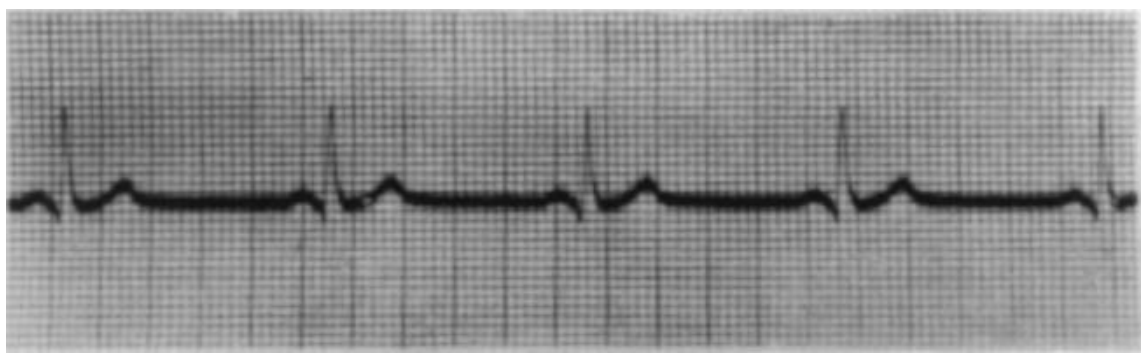

**Fig. 2: The first human ECG.**

The instruments that records such vibrations is called electrical cardiography (E.C.G) (figure 3) which is discovered by the scientist Willem Einthoven in 1902.

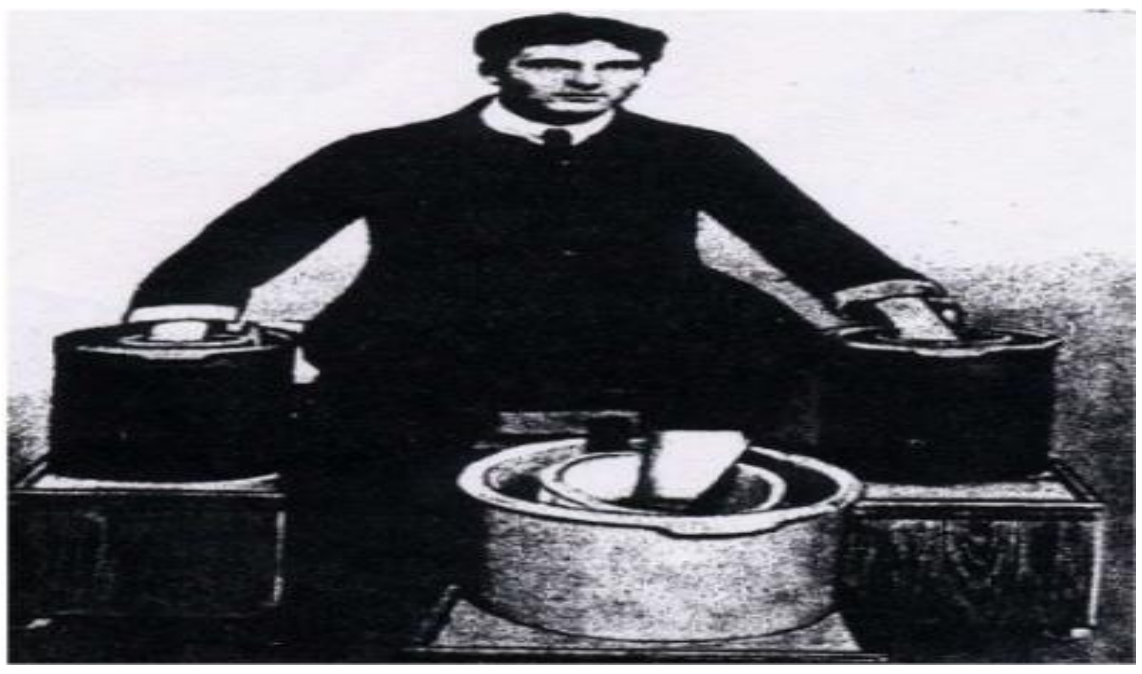

**Fig. 3: Depicting W. Einthoven recording his first ECG in 1902 by placing limbs in buckets of conducting solution.**

It translates the vibrations into electrical impulse which are then recorded (see the modern ECG in figure 4).

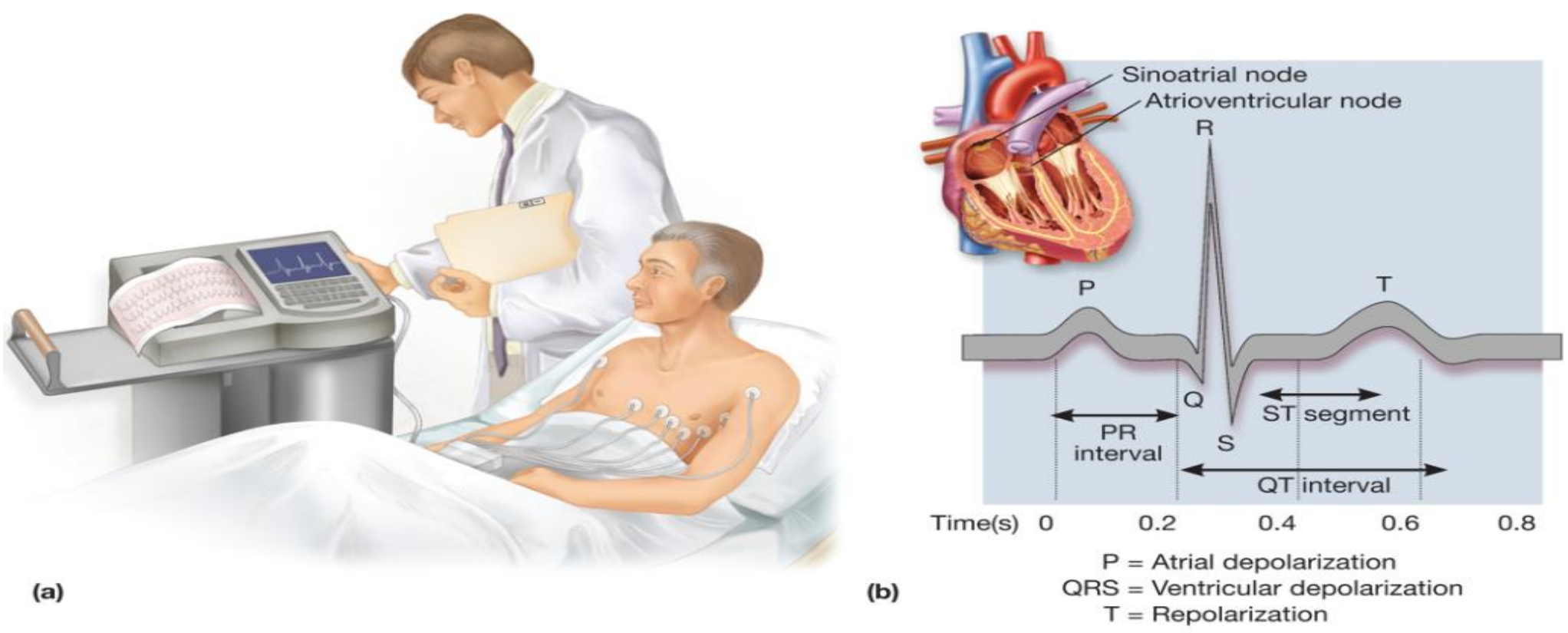

**Fig. 4: Modern ECG with sino-vibration graph.**

It is interesting to transport the heart vibrations into mechanical vibrations instead of translating these vibrations into electrical impulse. This can be done in the following manner. Now, suppose that a person rests on horizontal table which has springs so that it can vibrate horizontally, then, due to the pumping of heart the table will undergo small vibrations, the frequency and magnitude of which will depend various parameter associated with the heart. Some important conclusions about the vibrations of heart can be drawn. Let $y$ denote the horizontal displacement of some specified points of the table (as example, on end) from fixed point location (such as a wall). Let $M$ denote the combine mass of the person and the portion of table which is set into motion.
If we assume that there is a dumping force proportional to an instantaneous velocity and restoring for proportional to the instantaneous displacement. Then, the differential equation describing the motion of the table is given by [13]:

$$M\frac{d^2y(t)}{dt^2} + \beta\frac{dy(t)}{dt} + \gamma y(t) = F(t) \quad (42)$$

Where $\beta$ and $\gamma$ are constant of proportionality and $F$ is the force the system due to the pumping equation of heart. Suppose that $m$ is the mass of blood pumped out of heart luring such vibrations and $z$ is the instantaneous center of mass of this e quantity of blood. Then, by Newton's low, we have

$$F(t) = m\frac{d^2z(t)}{dt^2} \quad (43)$$

since $F(t)$ is the force which control the blood, then

$$F(t) = u(t) \quad (44)$$

thus, the dynamical system of the cardiography model becomes

$$M\frac{d^2y(t)}{dt^2} + \beta\frac{dy(t)}{dt} + \gamma y(t) = u(t) \quad (45)$$

**4. 2. The mathematical approach**

We use state space analysis to describe the physical dynamical system given by the following system

$$M\ddot{y}(t) + \beta\dot{y}(t) + \gamma y(t) = u(t) \quad (46)$$

The dynamical system (46) can be transform to the following state system given by the form

$$y = x_1$$
$$\dot{y} = x_2$$

By deriving the above equations implies that

$$\dot{x}_1 = x_2$$
$$\dot{x}_2 = \ddot{y} = -\frac{\gamma}{M}x_1 - \frac{\beta}{M}x_2 + u(t) \quad (47)$$

The dynamical system (4.4) given by matrix form as following

$$\dot{x}(t) = \begin{bmatrix} 0 & 1 \\ \frac{-\gamma}{M} & \frac{-\beta}{M} \end{bmatrix}\begin{bmatrix} x_1(t) \\ x_2(t) \end{bmatrix} + \begin{bmatrix} 0 \\ 1 \end{bmatrix}u(t) \quad (48)$$

Augmented with output function

$$y(t)\,] = [0 \quad 1]x(t) \quad (49)$$

**4. 3. The method**

We can use theorem 3.3 to prove that, the linear dynamical controlled systems $(48) - (49)$ are completely observable over $[0, T]$. Now we have.

$$\begin{bmatrix} \dot{x}_1(t) \\ \dot{x}_2(t) \end{bmatrix} = \begin{bmatrix} 0 & 1 \\ \frac{-\gamma}{M} & \frac{-\beta}{M} \end{bmatrix}\begin{bmatrix} x_1(t) \\ x_2(t) \end{bmatrix} + \begin{bmatrix} 0 \\ 1 \end{bmatrix}u(t)$$

augmented with output function

$$y(t) = [0 \quad 1]x(t)$$

we know that

$$A = \begin{bmatrix} 0 & 1 \\ \frac{-\gamma}{M} & \frac{-\beta}{M} \end{bmatrix}$$

and the conjugate transpose of $A$ is given by

$$A^* = \begin{bmatrix} 0 & \frac{-\gamma}{M} \\ 1 & \frac{-\beta}{M} \end{bmatrix}$$

thus,

$$C = [0 \quad 1]$$

and the conjugate transpose of $C$ is given by

$$C^* = \begin{bmatrix} 0 \\ 1 \end{bmatrix}$$

therefore, the matrix

$$[C^*, A^*C^*] = \begin{bmatrix} 0 & \frac{-\gamma}{M} \\ 1 & \frac{-\beta}{M} \end{bmatrix}$$

Since the determinant of ,

$$[C^*, A^*C^*] \neq 0$$

Then, we have

$$rank\,[C^*, A^*C^*] = 2 = n \text{ [3].}$$

Consequently, the system (48) together with the (49) is completely observable over $[0, T]$ ■.

## Conclusion

We have been presented some definitions and characterizations related to control system analysis in finite dimensional . More precisely, the observability problem of electrocardiography model has been studied and anlysis. Then, the existence of sufficient conditions which described the observability notion in linear dynamical systems are discussed and proved. Thus, we show that this physical model is completely observable to zero over finite time interval $t\epsilon\ [0, T]$. Many problem still opened for the future work, one can study the possibility of extending these results to the case of distributed parameter systems analysis, where the dimensional is infinite as in ref.s [15-19].

**Acknowledgements.** My thanks in advance to the editors and experts for considering this paper to publish in this esteemed journal. The author appreciate the time and effort in reviewing the manuscript and greatly value the assistant as reviewer for this paper.

# الشروط الكافية للقابلية على المشاهدة في مسألة راسمة القلب الكهربائية

رحيم احمد منصور الصفوري

*قسم الرياضيات ، كلية التربية للعلوم الصرفة ، جامعة تكريت ، تكريت ، العراق*

**E-mail:** saphory@hotmail.com

## الملخص

الهدف من البحث هو دراسة مسالة تخطيط القلب الكهربائية (Electrocardiography problem). وعلية سوف ننشأ منظومة فضاء الحالة كنموذج من النماذج الرياضية. علاوة على ذلك تم تقديم بعض التعاريف والنتائج التي توصف بعض المفاهيم في تحليل انظمة السيطرة الخطية التي تتعلق بتلك المسالة. وبشكل ادق، الشروط الكافية التي توصف مفهوم القابلية على المشاهدة في انظمة السيطرة الحركية الخطية قدمت ونوقشت. اخيرا، برهننا ان مسالة تخطيط القلب الكهربائية انها نموذج من الأنظمة القابلة للمشاهدة وعلى فتره زمنية منتهية مقدارها $t\epsilon\ [0, T]$.